\documentclass[12pt]{iopart}
\usepackage{graphicx}
\usepackage[sorting = none, backend = biber, style=numeric-comp]{biblatex}
\usepackage[justification=justified]{caption}
\usepackage{comment}
\usepackage[hidelinks]{hyperref}

\begin{document}
\title[Tutorial: How to build and control an all-fiber wavefront modulator]{Tutorial: How to build and control an all-fiber wavefront modulator using mechanical perturbations}


\author{Ronen Shekel$^{1*}$, Kfir Sulimany$^1$, Shachar Resisi$^1$, Zohar Finkelstein$^1$, Ohad Lib$^1$, Sébastien M. Popoff$^2$ and Yaron Bromberg$^{1\dagger}$}

\address{$^1$Racah Institute of Physics, The Hebrew University of Jerusalem, Jerusalem 91904, Israel}
\address{$^2$Institut Langevin, ESPCI Paris, PSL University, CNRS, France}
\ead{$^*$ronen.shekel@mail.huji.ac.il}
\ead{$^\dagger$yaron.bromberg@mail.huji.ac.il}

\vspace{10pt}

\begin{abstract}
Multimode optical fibers support the dense, low-loss transmission of many spatial modes, making them attractive for technologies such as communications and imaging. However, information propagating through multimode fibers is scrambled, due to modal dispersion and mode mixing. This is usually rectified using wavefront shaping techniques with devices such as spatial light modulators. Recently, we demonstrated an all-fiber system for controlling light propagation inside multimode fibers using mechanical perturbations, called the fiber piano. In this tutorial we explain the design considerations and experimental methods needed to build a fiber piano, and review applications where fiber pianos have been used. 
\end{abstract}

\vspace{2pc}

%
%
%

\section{Introduction}
In the past several decades, there has been an ongoing effort to utilize multimode optical fibers (MMFs) in photonic applications such as space-division multiplexing for optical fiber communication \cite{berdague1982mode, Richardson2013, Ploschner2015, puttnam2021space}, nonlinear optics \cite{baldeck1987observation, mafi2012pulse,wright2015controllable, krupa2017spatial, tzang2018adaptive,krupa2019multimode, teugin2020controlling,pourbeyram2022direct} and imaging \cite{vcivzmar2012exploiting,choi2012scanner,gu2015design,caravaca2017single,borhani2018learning,leite2021observing,lee2022confocal, stibuurek2023110, wen2023single}. More recently, applications for MMFs have expanded to quantum optics \cite{defienne2016two,valencia2020unscrambling,leedumrongwatthanakun2020programmable,cao2020distribution,sulimany2022all} and fiber lasers \cite{wright2017spatiotemporal,teugin2019spatiotemporal,haig2022multimode, sulimany2022soliton, fu2023recent}. These numerous applications enjoy the ability of MMFs to transport many spatial modes with low loss, while having a low cost and a small footprint. 

The main challenge of MMF-based technologies is that the information propagating through an MMF is distorted by inter-modal interference, modal dispersion, and mode coupling. The information delivered by an MMF is therefore often scrambled, yielding at its output complex spatiotemporal speckle patterns. 

One of the most promising approaches for unscrambling the information transmitted by MMFs is by shaping the optical wavefront at the proximal end of the fiber, to get a desired output at its distal end. Demonstrations include compensation of modal dispersion \cite{Shen2005, Alon2014,Wen2016}, spatial focusing at the distal end \cite{DiLeonardo2011,Papadopoulos2012, Caravaca-Aguirre2013,Papadopoulos2013,BoonzajerFlaes2018}, and delivering images \cite{Cizmar2012, Bianchi2012, Choi2012} or an orthogonal set of modes \cite{Carpenter:14, Carpenter2015} through the fiber. In wavefront shaping, this is typically done by controlling the incident wavefront using spatial light modulators (SLMs) or digital micromirror devices (DMDs). 

Recently, we demonstrated a new method for controlling light at the output of MMFs, which does not rely on shaping the incident light, but rather on inducing mechanical perturbations along the fiber. Such precisely controlled bends of the fiber, which are routinely used for polarization control in fibers \cite{ulrich1979polarization}, have only recently been proposed for controlling light in multimode optical fibers \cite{Golubchik2015, regelman2016method}. Inspired by these works, we apply computer-controlled bends at multiple positions along the fiber, in a configuration we call the \textit{fiber piano} \cite{resisi2020wavefront, resisi2021image, finkelstein2023spectral, shekel2023shaping}.

\begin{figure}
    \centering
    \includegraphics[width=\columnwidth]{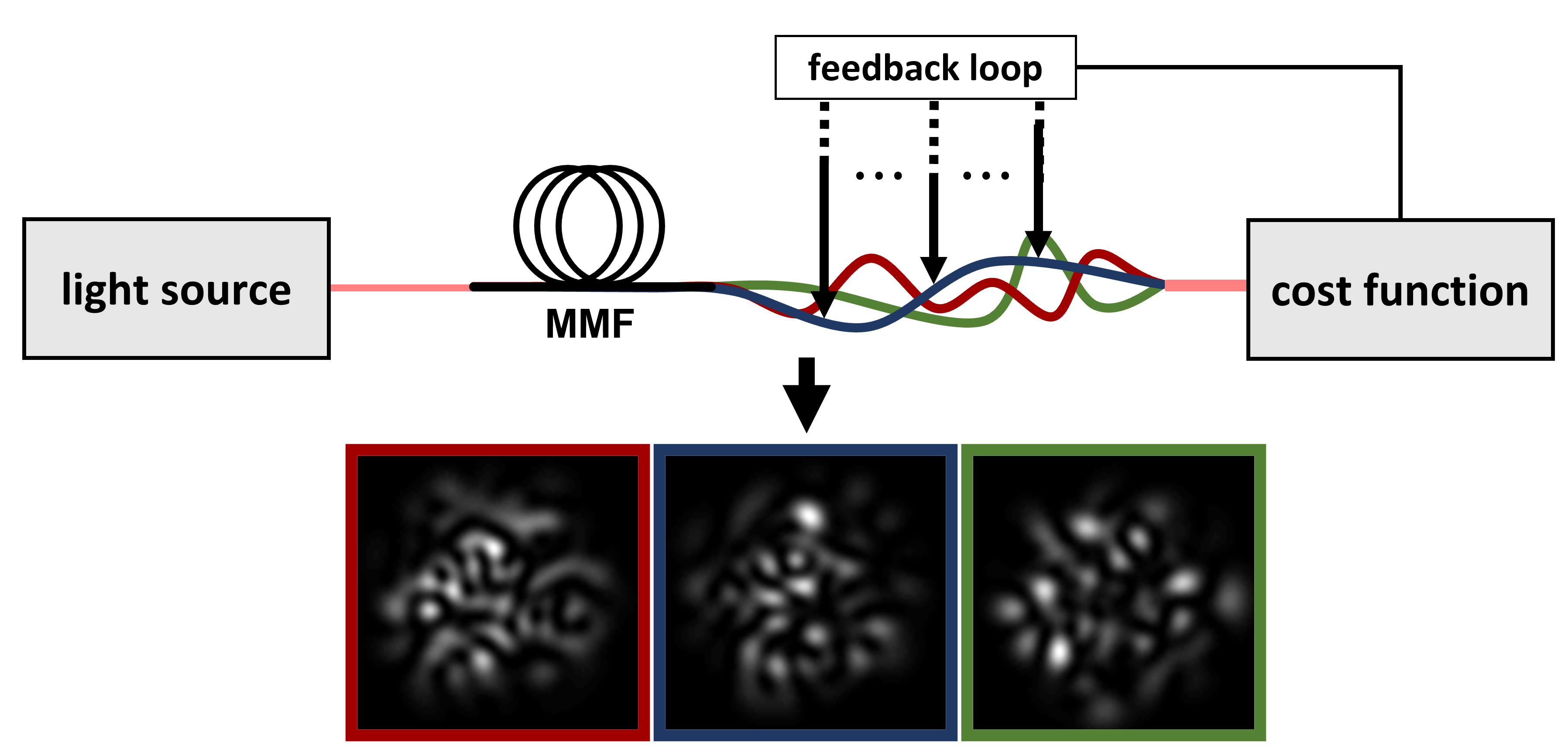}
    \caption{The working concept of the \textit{fiber piano}. Mechanical bends are applied along the multimode fiber. Different bends (depicted by red, blue, and green conformations of the fiber), yield different output patterns at the distal end of the fiber. To obtain a desired pattern, an optimization algorithm is employed to find the configuration of bends that optimizes a cost function of the output field.}
    \label{fig:concept}
\end{figure}

The bends we generate change the local propagation constants of the fiber’s guided modes and induce mode coupling \cite{taylor1984bending}, thus changing the fiber’s transmission matrix (TM). Specifically, as shown in Fig. \ref{fig:concept}, different bend configurations yield different spatial speckle patterns at the distal end of the fiber. 

We can therefore employ an optimization algorithm to search the space of bend configurations, to obtain a desired field distribution at the output of the fiber, without modifying the incident wavefront. Since in this approach the input field is fixed, it does not require an SLM or any other free-space component. In addition, in this approach, we shape the TM of the MMF, not the incident wavefront. Since the TM of a fiber that supports $N$ guided modes is determined by $\approx N^2$ complex parameters, the fiber piano provides access to many more degrees of control compared to when shaping the incident wavefront. 

In this tutorial, we provide in detail our experimental methods applied for building and using a fiber piano. In particular, we provide details on the actuators we use and how we place them along the fiber, as well as considerations in estimating the needed number of actuators. We then discuss considerations in choosing a suitable fiber and optimization algorithm. We also discuss several ways in which one may simulate a fiber piano. We end this technical section with details on how we define the enhancement in different applications. Finally, we briefly survey the different applications to which the fiber piano has been applied. In \cite{code_repo} we provide an example for how to simulate a fiber piano, and also provide raw data from several experiments.

\section{Experimental methods}

\subsection{The actuators}
To apply the macro bends along the fiber we use an array of piezoelectric actuators. Here, we discuss the design considerations of the fiber piano and its alignment, while detailed specifications of our actuators are given in the supplementary information. 

\subsubsection{Placement of actuators}
\begin{figure}
    \centering
    \includegraphics[width=\columnwidth]{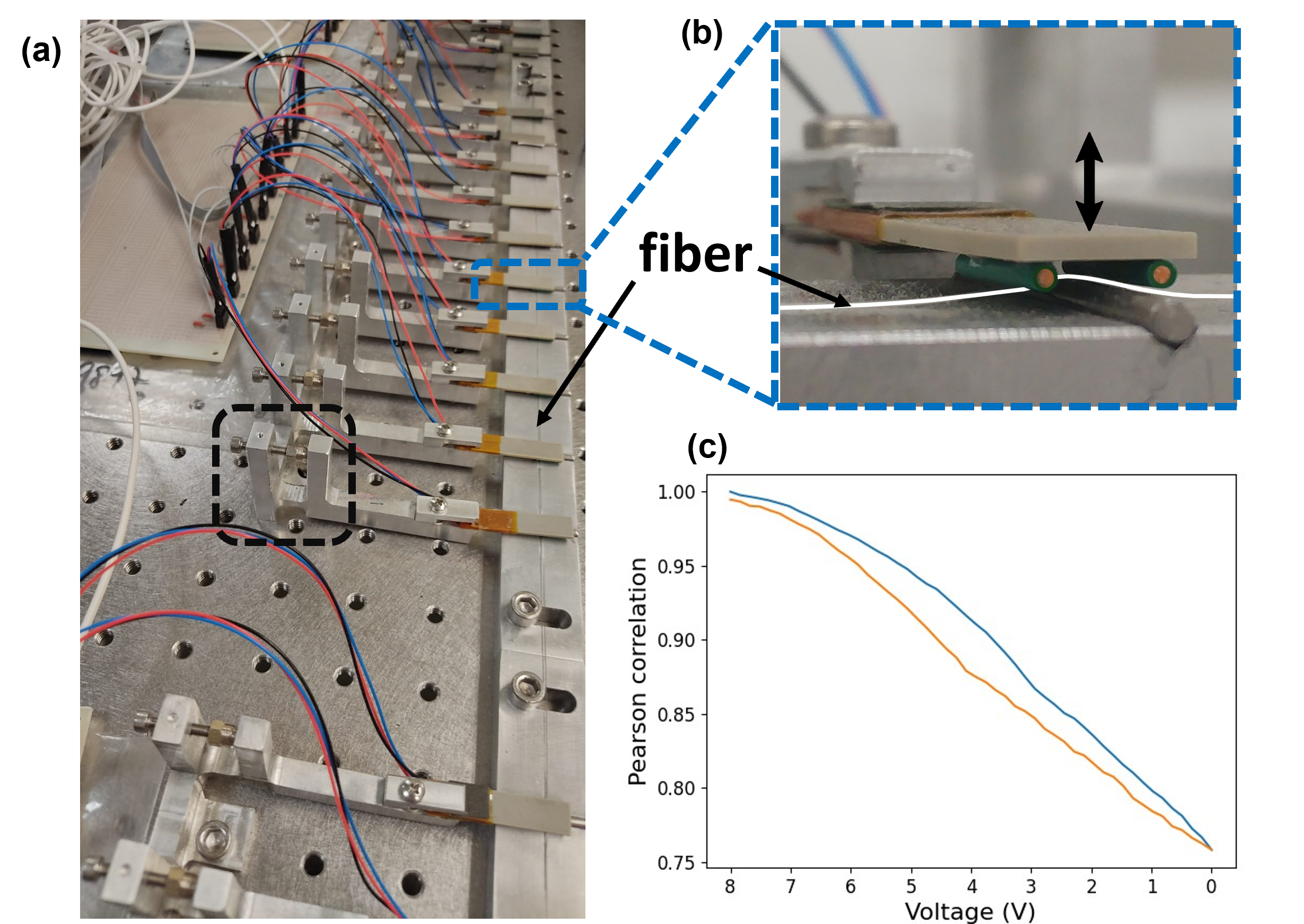}
    \caption{(a) A group of piezoelectric actuators used in the fiber piano. The black dashed square shows the mechanics used to align the initial height of the actuator above the fiber. (b) Zoom-in on a single actuator. The actuator moves up and down, generating a three-point bend of the fiber. The distance between the two rods of the actuator is $\approx 5$mm, and the actuator has a maximal travel of a few hundreds of microns, inducing curvatures of a few cm. For better visibility, we highlight the fiber by a white tracing. (c) A sample measurement of the effect of a single actuator, used to calibrate the height of an actuator above the fiber. Note that in our actuators the maximum voltage induces the smallest bend to the fiber, as our actuators bend upwards. See further in the supplementary information.} 
    \label{fig:actuators}
\end{figure}

In order for the spatial field distribution to significantly change upon propagation between two adjacent actuators, the distance between them must be larger than $d^2/\lambda$, for d the core’s diameter and $\lambda$ the wavelength. After this propagation length modes with two adjacent propagation constants will accumulate a relative phase of $\approx 2\pi$. In our case, the actuators were set 1.5-3 cm apart (see Fig. \ref{fig:actuators} (a)). 

The height of each actuator above the fiber is fixed such that when no bend was applied, the two small rods attached to the actuator (see Fig. \ref{fig:actuators} (b)) merely touch the fiber. To find this height and to estimate the voltage range to apply on the actuators, we send a laser through the fiber and output it to a camera. When no voltage is applied, we acquire an initial speckle pattern. We then scan the voltage on a single actuator, going from the smallest to the largest bend on the fiber, and back, recording the speckle patterns during the scan. We calculate the Pearson correlation (PCC) between these output speckles and the initial pattern, and choose the maximal voltage such that the correlation decreases to $\approx 80\%$ at the maximal bend, and returns to $\approx1$ when removing the bend. This corresponds to curvatures with radii of a few cm. 
The initial height, corresponding to the point of contact with the fiber, is found by making sure that a small voltage does induce some change in the PCC. 
The choice of a maximal decorrelation down to $\approx 80\%$ is chosen empirically; a bend too strong would increase the loss \cite{taylor1984bending, schermer2007improved}, which for the maximal deformation typically induces a few percent intensity attenuation per actuator. When the total power is very low, scanning the voltage and recording the speckle patterns could be very slow. In this case we merely switch the voltage between minimal and maximal bends and aim for a significant decrease in the PCC.

This process is repeated for each of the actuators. An example of such a measurement is depicted in Fig. \ref{fig:actuators}(c). Note that the correlation between patterns recorded when the same voltage is applied to the actuator is not perfect and shows some hysteresis. This is further discussed in the supplementary information.

\subsubsection{Number of actuators \label{num_actuators}}
In our experiments, we typically use a few tens of actuators. However, the number of actuators required for a fiber piano depends on the application for which it is used. We first consider the task of spatial focusing at the distal end of the fiber. In the regime where the number of modes excited in the fiber $N_{modes}$ is much larger than the number of actuators $N_{act}$ it has been shown both in simulation and experimentally \cite{resisi2020wavefront}, that the enhancement of the focus increases linearly with the number of actuators, with a slope of $\alpha\approx0.7$. An optimal enhancement may be achieved when $N_{act}\approx N_{modes}$, therefore for a task such as focusing, there would be redundancy in the effect of the actuators if using more actuators. In fact, adding additional actuators might deteriorate performance due to the additional losses. 

To take both effects into account, we can multiply the empirically found enhancement of $\alpha\cdot N_{act}+1$ by a factor of $t^{N_{act}}$, where $\alpha$ is the enhancement linear pre-factor, $t<1$ is the transmission of a single actuator, and the $+1$ accounts for the unity enhancement when no actuators are used. With this model the ideal value for $N_{act}$ is $N_{act}=1/\ln{\frac{1}{t}} - 1/\alpha$. For $t=0.97$ ($0.98$) we get an optimal $N_{act}\approx 31$ ($48$). Note that the parameters $t$ and $\alpha$ may be controlled to some extent by different choices of the amount of travel of the actuators, or by other actuator designs.

For applications that require shaping of multiple incident wavefronts simultaneously, having $N_{act}> N_{modes}$ may be required. Unlike shaping using an SLM, where the shaping is performed on the light incident on the MMF, the fiber piano allows tayloring the transmission matrix of the MMF itself. Thus, while the degrees of control available using an SLM are the $N$ complex amplitudes of the input modes, with the fiber piano one may control the $\approx N^2$ amplitudes of the transmission matrix of the MMF. This higher degree of control conceptually allows one to perform more demanding tasks, such as implementing arbitrary unitary transformations between the input and output modes of the MMF. This is similar to the freedom given by methods such as multi-plane light conversion (MPLC) \cite{morizur2010programmable, labroille2014efficient}, where it has been shown that $N^2$  phase shifters may be programmed to perform arbitrary transformations, even when they are not independent \cite{tanomura2022scalable, taguchi2023iterative}. 

Performing such tasks with the fiber piano is currently challenging, however, since the optimal configuration of the actuators is currently reached by a "blind" optimization algorithm, which may take very long to converge. This is compared to the more sophisticated methods applied in MPLC, such as the phase matching algorithm \cite{fontaine2019laguerre}. In order to apply such algorithms to the fiber piano, we first must attain a more accurate model of the perturbations it induces. This can be achieved by precise measurements of the changes each actuator induces on the transmission matrix \cite{matthes2021learning}. 

A second, conceptual, challenge is to identify the effective amount of degrees of control provided by a single actuator. While for phase modulation the maximum effect per pixel is applying a $2\pi$ modulation, in our case the effect is not limited. Supposing that a bend induces some mode coupling between nearest neighbors, applying a further bend will further propagate the mode coupling to next-nearest neighbors. The effective amount of control should thus be connected to the decorrelation induced by its maximal voltage, or to $\alpha$ the enhancement parameter discussed above.

\subsection{The fiber}
Choosing a suitable fiber may prove critical for specific applications, and several trade-offs must be considered. A simple example is the core size: while a larger core allows easier coupling into the fiber, it will support more modes, which could require more actuators and a longer optimization.

Another trade-off has to do with the fiber type, e.g. a step-index (SI) or a graded index (GRIN) fiber. The modes of SI fibers have larger modal areas and more closely spaced propagation constants for the same numerical aperture, which leads to increased mode coupling and higher sensitivity to bending \cite{BoonzajerFlaes2018, qiu2024spectral}. Thus, using an SI fiber allows more control per actuator, which may decrease the required amount of actuators. However, SI fibers have a much smaller spectral correlation width than GRIN fibers \cite{Pikalek:19}. This makes spatial shaping of a broadband signal impractical since different wavelengths will produce different speckle patterns that will sum incoherently. In applications requiring a broadband operation, for example, shaping photons generated via spontaneous parametric down conversion (SPDC) \cite{shekel2023shaping}, we used an $\approx 1.8$m long OM2 (GRIN) fiber. The spectral correlation width of this fiber was $\approx 2.8$nm, allowing us to use bandpass filters with a full-width half max (FWHM) of $3$nm. In \cite{resisi2021image}, we used a HeNe laser which has a very small linewidth, allowing us to use an $\approx 1.5$m long SI fiber, even though it has a much smaller spectral correlation width \cite{Pikalek:19}. 

While the length of the fiber does not directly affect the optimization process, using a long fiber will make the spectral correlation width narrower, and may also degrade the general stability of the system.

A related issue is that of polarization. In many cases strong mixing occurs in the fiber, and different polarizations exhibit different speckle patterns. Since the two polarizations sum incoherently, the output non-polarized speckle will have a lower contrast than that of a polarized output. To solve this, we typically separate the two polarizations using a polarizing beam splitter (PBS) or a Wollaston prism. 

\subsection{Optimization algorithms}
As described above, the fiber piano approach relies on solving optimization problems. These problems use a feedback loop - at each iteration, the output field is recorded and evaluated according to some cost function. This cost function is then given to the optimization algorithm, which tries to minimize it by changing the configuration of bends that are applied to the fiber segments. Since the propagation of the light in each curved segment depends on the deformations induced by all the actuators prior to it, the ideal way to deal with such an optimization problem is using a nonlinear optimization algorithm.

In our works, we use a particle swarm optimization (PSO) algorithm \cite{resisi2020wavefront, shekel2023shaping} or simulated annealing \cite{finkelstein2023spectral}. We also found that a genetic algorithm (implemented in Matlab) provides performance comparable to these two algorithms. We have also seen in simulations (unpublished) that hybridizing several algorithms could be beneficial: we start the optimization with an algorithm known to converge quickly, and once it begins to saturate, we switch to an algorithm that generally shows better enhancement. 

The optimization algorithm we chose in \cite{resisi2020wavefront, shekel2023shaping} is PSO, a genetic algorithm. It randomly initializes a population of points (referred to as particles) in an $N_{act}$-dimensional search space, representing the voltages that are assigned to the $N_{act}$ actuators. These positions are iteratively improved according to their local and global memory from previous iterations. Its stochastic nature helps avoid local extrema in non-convex problems.

The optimization algorithm we chose in \cite{finkelstein2023spectral} is the simulated annealing (SA) algorithm \cite{rutenbar1989simulated}. To decrease the probability of converging to local minima at each iteration, we added with some probability $p$ a weak perturbation to the optimal configuration of actuators, and increased $p$ as the temperature decreased.

In both cases, with a few tens of actuators, the optimization process typically took several hundred to a few thousand iterations. This translated to a timescale of tens of minutes, due to slow control electronics. 

All the algorithms mentioned above are "blind", using no prior knowledge about the actual effect of the piano perturbations. Measuring the exact effect of the actuators might help with performing the optimization in a more efficient manner.

\subsection{Simulation}
Since our system is linear in the optical field, it is natural to describe it with a matrix formalism. It is convenient to model the fiber with consecutive segments of straight and curved fibers. The curved segments model the effect of actuators, where the curvature of the segment is proportional to the travel of the actuator. The straight segments describe the propagation between the actuators. The transmission matrix (TM) for curved and straight segments is estimated numerically, and the complete fiber TM is calculated by multiplying the TMs of all segments. In each iteration of the optimization process, the algorithm chooses the bend configuration to apply by selecting the set of corresponding TMs such that the cost function is optimized.

Simulating a straight fiber is straightforward: First, we compute the mode profiles and propagation constant of the ideal fiber. The TM in the fiber mode basis then corresponds to a diagonal matrix whose elements represent the phase accumulations $e^{i\beta_mL}$, where $L$ is the segment length and $\beta_m$ is the propagation constant of the $m^{th}$ mode. For the bent segment, there are several approaches to simulate the TM. The simplest yet least accurate approach is to assume the bending induces mode mixing that can be represented by a random unitary matrix. Different bends can then be simulated by taking different random matrices. The strength and nature of the mode mixing can be set by applying constraints to the unitary matrices. For example, a block diagonal matrix can model strong mixing only within the degenerate mode groups in graded index fibers, a band matrix can model nearest (or n-nearest) neighbor coupling between adjacent modes. 

A second approach to model the TM of a curved segment is to approximate the bends by a circular arc, with a curvature defined by the travel of the actuator. Then, the guided modes and propagation constants could be found using numerical modules such as \cite{pymmf} or \cite{veettikazhy2021bpm}, which solve the scalar Helmholtz equation under the weakly guiding approximation \cite{OKAMOTO2006}. Then, the maximal and minimal radii of curvatures applied in the experiment should be estimated from the experimental system, and different curvatures could represent different perturbations of an actuator. In \cite{resisi2020wavefront} this approach was used, in combination with the first approach: each actuator was modeled as a product of two matrices - one emulating a circular arc, and the second a random block diagonal matrix, coupling the modes within the same degenerate mode space. Without introducing the random coupling matrices we were unable to achieve focusing in the simulation. 

A third approach for modeling the effect of a bent fiber, is to measure its TM. Doing this requires accurate measurements of the TM in the fiber-mode basis, for each bending radius. Indeed, such measurements have been made in \cite{matthes2021learning}. The advantage of such a simulation is that it also captures polarization-dependent effects and polarization mixing. In \cite{code_repo} we provide a simulation of the fiber piano based on this approach, and the TM measurements performed in \cite{matthes2021learning} for a slightly different type of mechanical perturbation.

\subsection{Enhancement}
In most applications, we use the fiber piano to optimize a cost function. In the context of focusing (see \ref{app:spatial} below), it is convenient to set the cost function as the enhancement factor $\eta$, defined as the total power in some region of interest (ROI), divided by the total power in the ROI before the optimization. Since the intensity in the ROI before optimization is very sensitive to the actuators' configuration, it is computed by averaging the output intensity over random configurations of the actuators. Similarly, in the context of spectral shaping (see \ref{app:spectral} below), we define the enhancement by the peak-to-background ratio in the measured spectra. In the spectral case there is no need for averaging over random actuator configurations and measuring the intensity of the exact same wavelength, since there is no envelope over the field of view that differentiates between different areas of the spectrum.

The enhancement is a good measure for applications in which the primary goal is to shape the spatial distribution at the output of the fiber, where the total power transmitted through the fiber is a secondary concern. However, in applications that require high transmittance through the fiber, one must consider the trade-off between the amount of loss the actuators induce and the enhancement they can achieve. In such a case a better cost function would be the total power in the ROI, with no normalization.

\section{Applications}
The fiber piano has been used for several applications, which we briefly survey below. 

\subsection{Spatial shaping \label{app:spatial}}
Our first demonstration using the fiber piano to control light at the output of an MMF was 
focusing the light at the output of the fiber to a single spot \cite{resisi2020wavefront}. We excited a subset of the fiber modes by weakly focusing an input HeNe laser on the proximal end of the fiber. Due to inter-modal interference and mode-mixing, at the output of the fiber, the modes interfere in a random manner, exhibiting a fully developed speckle pattern. 

\begin{figure}[h]
    \begin{centering}
    \includegraphics[width=0.85\linewidth]{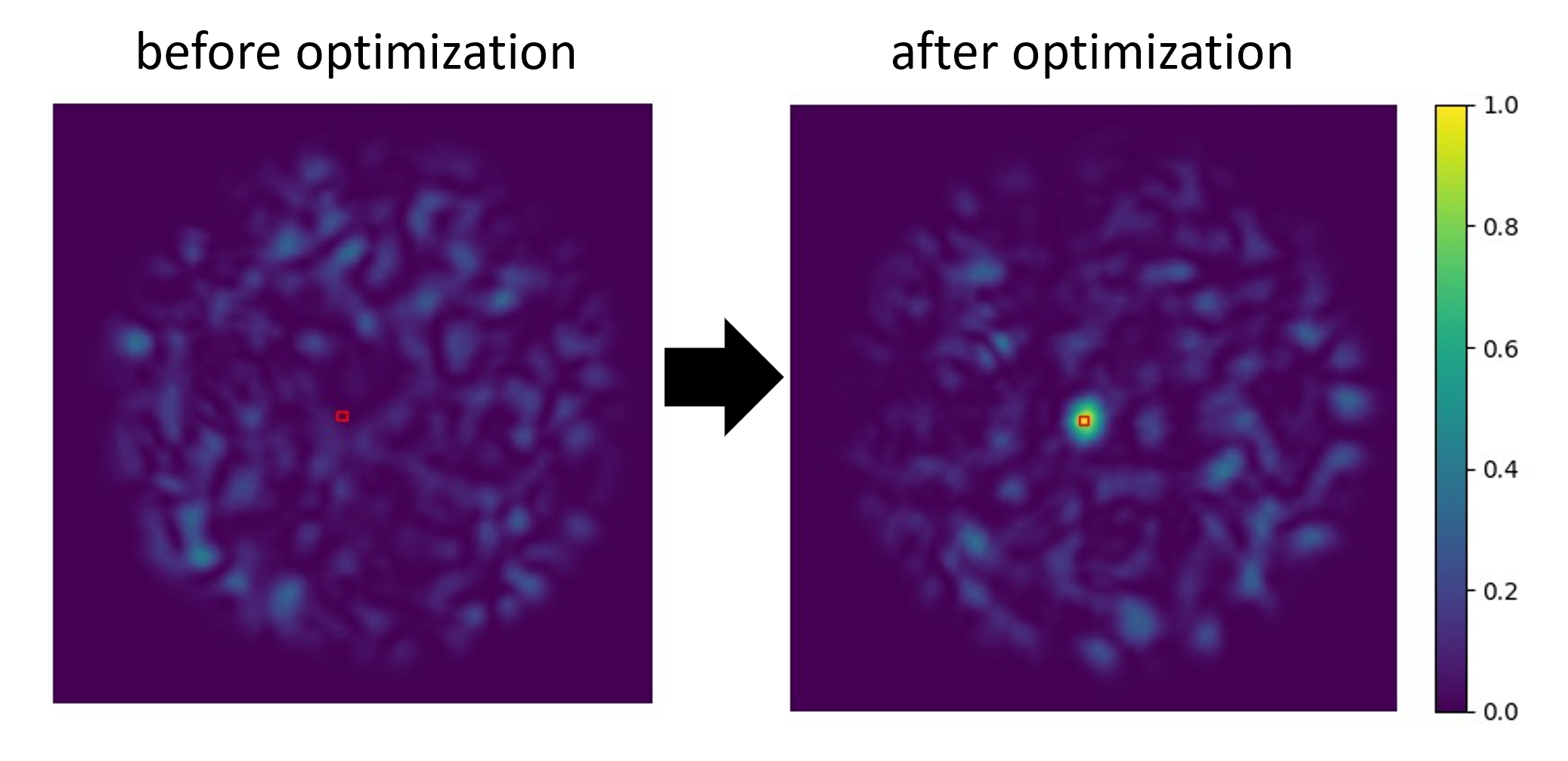}
    \par\end{centering}
    \caption{An example of focusing coherent light at the output of an MMF using the fiber piano, showing an enhancement of $\eta\approx18$. (a) The speckle pattern at the fiber output before the optimization process. (b) The speckle pattern at the fiber output after the optimization process. The red square depicts the region of interest we defined, which in this case was slightly smaller than the size of a speckle grain.}
    \label{fig:focusing}
\end{figure}

We defined a small ROI surrounding a chosen position, roughly the size of a single speckle grain, and run the PSO algorithm to maximize the total intensity in that area. Fig. \ref{fig:focusing} depicts the output speckle pattern of the horizontal polarization before (a) and after (b) the optimization, using 37 actuators. The enhanced speckle grain is clearly visible, and has a much higher intensity than its surroundings, corresponding to an enhancement of $\eta \approx 18$.

We have also shown in \cite{resisi2020wavefront} that we can tailor the output patterns in a few-mode fiber, where the number of fiber modes is comparable to the number of actuators. Specifically, we converted a superposition of guided modes to one of the linearly polarized (LP) modes supported by the fiber. To this end, we utilized the PSO optimization algorithm to find the configuration of actuators that maximizes the overlap between the output intensity pattern and the desired LP mode, which were computed numerically. For example, we demonstrated conversion of horizontally polarized LP11 mode to a superposition of a horizontally polarized LP01 mode and a vertically polarized LP11 mode. The Pearson correlation between the target and final patterns, in this example, was 0.93. Similar results were obtained when we ran the optimization using as few as 12 actuators, with a reduction of only a few percent in the correlation between the target and the final pattern. The performance deteriorates when using less than 12 actuators, as the number of actuators becomes comparable with the number of guided modes.

\subsubsection{Shaping quantum light \label{app:quantum}}

We have further shown that the fiber piano may also be used for spatial shaping of quantum light \cite{shekel2023shaping}. A photon is a single excitation of an optical mode, and when it propagates through a multimode fiber, its spatial distribution will be scrambled, similarly to a laser occupying the same mode. In \cite{shekel2023shaping}, we used SPDC to generate a heralded single photon and send it through an MMF. We used a single photon detector for feedback, and performed an optimization process using the fiber piano to localize the spatial distribution of the photon at the output of the fiber. 

Next, we sent spatially entangled photons generated via SPDC \cite{walborn2010spatial} through the MMF. Similar to classical light, which is scrambled in the fiber and  exhibits a speckle pattern, the spatial correlations between the two photons get scrambled, yielding a two-photon speckle \cite{peeters2010observation}. We use the correlations between the two photons for feedback, and perform an optimization process using the fiber piano to localize the spatial correlations. 

We note that in this experiment, we used a GRIN fiber and not an SI, even though it is less sensitive to mechanical perturbations. This is because a GRIN fiber has a much larger spectral correlation width \cite{Pikalek:19}, and the SPDC signal can be tens of nm wide. Even using a graded index fiber, we needed to use narrow-band spectral filters (3nm FWHM), further reducing the inherently weak SPDC signal. 

\subsubsection{Coupling into a single mode fiber \label{app:to_smf}}
A simple application for the focusing experiments is to couple light entering an MMF into a single-mode fiber (SMF). In many cases, it is desirable to use SMFs due to their lower dispersion and polarization mixing, easier integration with detectors, and better components and infrastructure. However, in applications where the incoming light is distorted, such as in free-space communications and imaging, coupling light directly into an SMF limits the collection efficiency due to the need to map the distorted incoming wavefront to the single guided spatial mode supported by the fiber. 

In \cite{shekel2023shaping}, we showed that we could use the MMF, together with a fiber piano, to efficiently funnel light into an SMF, getting the best of both worlds. The collection from free space is done by the MMF, which, thanks to its large core and multiple guided modes allows higher collection efficiencies than a single mode fiber. Using the fiber piano, the light is then coupled into the single mode fiber. Comparing to traditional adaptive optics systems, we consider a collection telescope for communication with a satellite. In such settings, it may be inconvenient to mount on the telescope a bulky adaptive optics system. In our approach, however, we do not need to add components on the telescope, and simply change the collecting fiber from an SMF to an MMF, with the fiber piano located in a more convenient position. We note that a possible disadvantage of adding an MMF to the collection system in communications, is the fact that it may add non-negligible mode-dependent polarization rotations, scrambling polarization-encoded information. Also, for information encoded in time, such as in telecommunication applications, dispersion in the fiber may limit the link speed. However, using a few-meter MMF should allow GHz rates of communication.

\subsection{Spectral shaping} \label{app:spectral}

The speckle pattern at the output of an MMF is wavelength-dependent, where the sensitivity to the wavelength grows with the length of the fiber. Using this feature, we utilized the fiber piano to realize a reconfigurable spectral filter \cite{finkelstein2023spectral}. To this end, we coupled a broadband fiber-coupled LED source to an SMF, yielding spatially coherent broadband light. We then spliced the SMF to a 1m step-index MMF. A subsequent splice to another SMF leads the light to a spectrometer. The SMF fiber acts as a spatial filter that selects one speckle channel of the speckle pattern generated at the output of the MMF. This speckle changes when the input wavelength changes, with a spectral decorrelation range inversely proportional to the dispersion. 

Using the fiber piano and the simulated annealing optimization algorithm, we find a bending configuration that maximizes the overlap between the output spectrum and a target spectrum. This is equivalent to a focusing experiment at the chosen frequency range. Because the speckle spectrally decorrelates, so does the focal spot~\cite{van2011frequency}, leading to a drastically lower transmission through the system for wavelengths outside the target spectrum. 

\begin{figure}[h]
\begin{centering}
\includegraphics[width=0.85\linewidth]{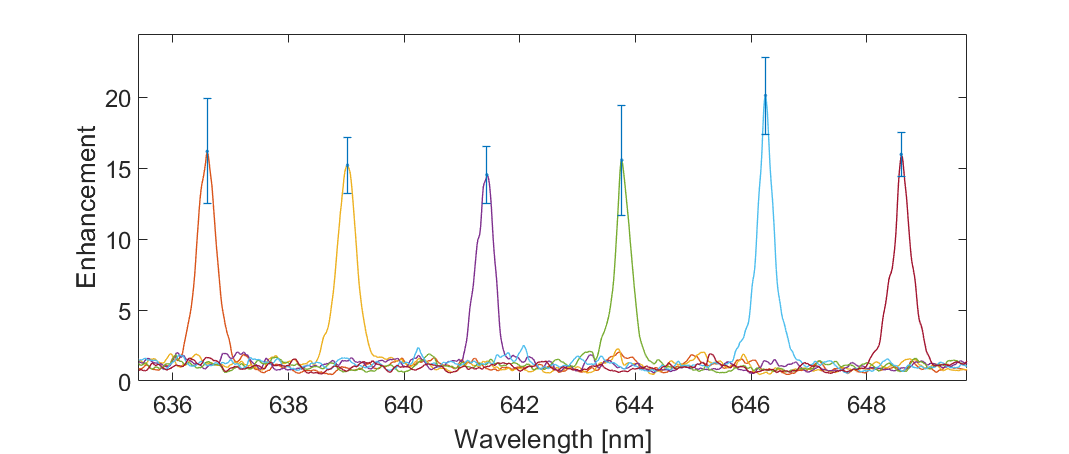}
\par\end{centering}
\caption{The measured transmission spectra at the output of a fiber piano spliced to a single mode fiber, exhibiting a tunable bandpass filter. Each curve represents an average of 6 optimized spectra to maximize the enhancement at the desired wavelength. We achieve a full-width-half-maximum of $\approx245$pm and enhancements in the range of 13 to 20.}
\label{fig:app_spectral}
\end{figure}

This system allows creating a versatile, tunable, spectral filter. For instance, we could optimize the piano to enhance the transmitted intensity at an arbitrary wavelength. In this way, we could demonstrate a tunable spectral filter (Fig. \ref{fig:app_spectral}). Similarly, a dual-band filter can be obtained by enhancing the intensity at two different wavelengths, simultaneously. The spectral resolution of the filters is determined by the resolution of the spectrometer and the spectral correlation width of the MMF, scaling as $1/L$ for MMFs with weak coupling between modes \cite{Redding:12, redding2013all, Cao:23}, where $L$ is the length of the fiber.

To create very narrow filters, one can use a long MMF to reduce its spectral correlation width. The resolution is then limited by the resolution of the spectrometer used in the optimization process. However, it is possible to replace the spectrometer and perform the optimization using a narrowband tunable laser. We used a tunable laser at telecom wavelengths that can tune the wavelength in the C-band with a spectral resolution of 1 pm and a camera that images the MMF output. We then used the fiber piano to focus on a single spot at a specific wavelength. After the optimization, we scanned the wavelength of the laser and observed the transmitted spectrum at that spot. 

The main advantage of using the fiber piano for spectral filtering is that its resolution is inversely proportional to the fiber length \cite{redding2012using, redding2014high}, allowing us to demonstrate filters with a 5 pm resolution using a 150m long fiber. Moreover, in contrast to grating-based reconfigurable filters, the resolution of the fiber piano is not limited by the operating wavelength range, opening the door for realizing high-resolution reconfigurable spectral filters operating over broad bandwidths. 

Another advantage is that, unlike wavelength-selective technologies, the resolution of the filter does not come at the cost of a large footprint. As we do not use a grating, we may coil a 100m fiber onto a 3-inch spool, and, in principle, create a 1 pm resolution spectrometer at 1500nm, surpassing even 1-meter-long bench-top grating spectrometers \cite{redding2014high}.

The main drawback of this system is its relatively low efficiency, defined by the output-to-input intensity ratio in the chosen spectral band, which was 0.06 in our experiments. This, however, can potentially be improved by using more actuators to achieve better enhancements, and by designing better actuators that have lower loss.

\subsection{Control of nonlinear processes in multimode fibers\label{app:nonlinear}}
The complex spatiotemporal dynamics and inter-modal interactions inherent to multimode fibers open vast possibilities in the realm of nonlinear wave propagation \cite{krupa2019multimode}. This field has fostered innovations in diverse areas such as fiber lasers \cite{wright2017spatiotemporal, teugin2019spatiotemporal, sulimany2022all}, frequency conversions \cite{demas2015intermodal, lopez2016visible, krupa2016observation, eftekhar2017versatile}, photon pair generation \cite{sulimany2022all, garay2023fiber}, and beam cleaning \cite{krupa2017spatial}. In recent years, there has been significant interest regarding the control of nonlinear phenomena within multimode fibers, achieved through wavefront shaping techniques \cite{wright2015controllable, tzang2018adaptive, eftekhar2019accelerated, deliancourt2019wavefront, teugin2020controlling}. These methods principally involved manipulating the wavefront in free space, before coupling into the fiber, typically using a spatial light modulator. However, this methodology does not permit modulation of the light while inside the non-linear propagation medium, or shaping of the nonlinear process itself. 

Qiu et al. have recently shown that one can resolve this challenge using a fiber piano \cite{qiu2024spectral, liu2024deep}. The control of the wavefront was achieved during nonlinear light propagation in multimode fibers using a fiber piano that was built using 3D printing and low-cost stepper motors. When applied to a standard silica SI fiber at high power levels, they demonstrated control of the propagating pulses across the temporal and spatial domains. Remarkably, they incorporated the fiber piano inside a multiphoton microscope, showing an efficient and highly adaptable two-photon and three-photon microscopy, for both fluorescent beads and label-free tissues.

\subsection{Learning macro bends}
All applications mentioned above had to do with optimization and wavefront-shaping. However, in \cite{resisi2021image}, the fiber piano was used in a different context, of sending images through an MMF. This, in general, is a challenging task, as the propagation through the MMF scrambles the image. To overcome this, a convolutional neural network (CNN) was trained on many different pictures, while the MMF was distorted in many different configurations using the fiber piano. When the training was completed, the CNN could reconstruct pictures sent through the fiber, also for new piano bending configurations that were not used in the training process, and even generalized to macro deformations on a larger length scale of a few cm. In this context, the piano was used simply to generate many different random mechanical configurations of the fiber, generating data to feed into the CNN.

\section{Outlook}
In this paper, we presented the fiber piano in detail, providing information on different aspects and considerations in building one and applications where it has been used. The fiber piano allows a robust and stable all-fiber operation for various tasks. Given further development, we believe it may be used to tailor a full transmission matrix of a multimode fiber, similar to what has been done with MPLC \cite{kupianskyi2023high} and with configurable circuits using MMFs together with SLMs \cite{matthes2019optical, leedumrongwatthanakun2020programmable, goel2024inverse}. 

We believe that with proper design and engineering, one could construct fiber pianos with lower losses and faster response times, at lower costs. This may be achieved either by using faster mechanical actuators, such as those used in deformable mirrors \cite{archer2016dynamic}, or by transferring the fiber-piano concept to other types of actuators, such as in-fiber acousto-optical or electro-optical modulators. Such developments may open the door for many more practical applications. 

\section*{Funding}
This research was supported by the Zuckerman STEM Leadership Program, the Israel Science Foundation (grant No. 2497/21 and 1268/16), the State of Lower Saxony, Hannover, Germany, and the ISF-NRF Singapore joint research program (Grant No. 3538/20). R.S. acknowledges the support of the HUJI center for nanoscience and nanotechnology, and of the ministry of innovation, science and technology, Israel. O.L. acknowledges the support of the Clore Scholars Programme of the Clore Israel Foundation. S.M.P. acknowledges the support of the French Agence Nationale pour la Recherche (Grant No. ANR-20-CE24-0016 MUPHTA and ANR-16-CE25-0008-01 MOLOTOF), of Labex WIFI (ANR-10-LABX-24, ANR-10-IDEX-0001-02 PSL*), of Laboratoire International Associé ImagiNano, and of the France's Centre National de la Recherche Scientifique (CNRS; France-Israel grant PRC1672)

\printbibliography 

\end{document}


\title[Supplementary information]{Supplementary information for Tutorial: How to build and control an all-fiber wavefront modulator using mechanical perturbations}


\author{Ronen Shekel$^{1*}$, Kfir Sulimany$^1$, Shachar Resisi$^1$, Zohar Finkelstein$^1$, Ohad Lib$^1$, Sébastien M. Popoff$^2$ and Yaron Bromberg$^{1\dagger}$}

\address{$^1$Racah Institute of Physics, The Hebrew University of Jerusalem, Jerusalem 91904, Israel}
\address{$^2$Institut Langevin, ESPCI Paris, PSL University, CNRS, France}
\ead{$^*$ronen.shekel@mail.huji.ac.il}
\ead{$^\dagger$yaron.bromberg@mail.huji.ac.il}

\vspace{10pt}

\vspace{2pc}
\noindent{\it Keywords}: multimode fibers, wavefront shaping, fiber piano

%
%
%

\section{Actuator specifications}
The actuators we use are CTS NAC2225-A01 piezoelectric plate benders, where each actuator has three input wires: ground, some constant voltage $V_{max}$, and a variable voltage. We supply the constant voltages using an ES150 DC voltage supplier \cite{DCsupply}. A computer-controlled digital-to-analog converter (DAC) \cite{DAC} together with an amplifier \cite{amplifier} supplies voltages between $0$ and $V_{max}$ to the different actuators in order to achieve different bending configurations. In our case, applying higher voltages bends the plate upwards, creating smaller bends on the fiber. These benders allow voltages up to $200V$, with a free stroke of $\pm375 \mu m$. 

The benders were placed along the fiber using a mechanical structure we designed, holding the actuators in place and allowing manual control over the height of the actuators above the fiber. We attach two small rods (made of standard electrical wires) to the bottom of the actuator, such that when the actuator bends down, a three-point contact is made on the fiber, inducing a bend (Fig 2 (b) in the main text). Other types of perturbations might prove better for different applications. For instance, applying twists might prove more effective to control polarization. 

The actuators were divided into four groups, in which the spacing between nearby actuators is set to be 1.5-3 cm. The groups were placed along different fiber segments, at distances that were arbitrarily chosen.

Other mechanical actuators and configurations could be suitable as well. For example, in \cite{qiu2024spectral}, a different implementation of a fiber piano was realized using stepper motors, as discussed in the applications section of the main text.

\subsection{Response Time}  
A critical parameter of the fiber piano is its response time, i.e. the time it takes the output intensity pattern to stabilize after an abrupt change is applied to actuators. This parameter ultimately determines the optimization speed. To measure the typical response time, we introduced abrupt changes to the voltages applied to a subset of the piezoelectric actuators, and recorded the speckle pattern obtained at the distal end of the fiber as a function of time.

\begin{figure*}[htb]
\begin{centering}
\includegraphics[width=0.85\linewidth]{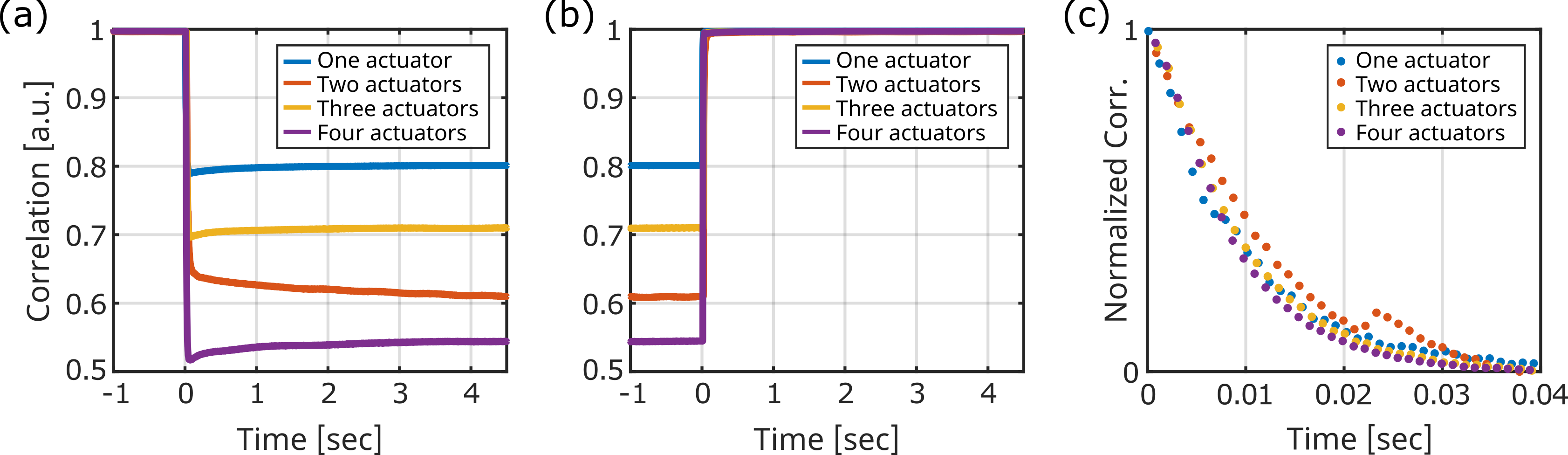}
\par\end{centering}
\caption{The 2D Pearson correlation coefficient of each frame with (a) the first and (b) the last of the acquired frames, when a configuration of actuators (the voltage which is applied on these piezos) is changed. (c) ‘zoom in’ on the abrupt change at time $t=0 \hspace{2pt}sec$ shows a response time of $\approx30 \hspace{2pt}ms$. Blue lines/dots show a configuration change of a single actuator, red for two actuators, yellow for three and purple for four. Adapted from \cite{resisi2020wavefront}.}
\label{fig:response}
\end{figure*}

We then calculated the two-dimensional Pearson correlation coefficient between each of the captured frames and the first frame. The measurements were repeated using different subsets of piezos. Examples of a few of these measurements, for subsets that include between one and four actuators, are shown in Fig. \ref{fig:response}(a). The abrupt voltage change causes a fast change in the recorded speckle pattern, yielding a sharp decrease in the computed correlation coefficient. As expected, the larger the subset of the piezos, the stronger the correlation drop. This sharp decrease results from the change in the actuators' configuration (the bend they pose) and manifests the significant change in the output intensity pattern. Fortunately, the response time remains practically the same when moving a single or multiple actuators simultaneously, allowing us to move many actuators simultaneously in each optimization step.

Once the actuator's position stabilized, the correlation stabilized on a lower value. To ensure that the patterns with lower correlation with the first frame are correlated with one another (thus ensuring that the plateau is not a result of the statistical properties of speckle patterns), we also calculated the 2D correlation coefficient of each frame from the final acquired frame. These results are shown in Fig. \ref{fig:response}(b) for the same groups of actuators. The high correlation after the configuration change indeed verifies that the speckle pattern did not change further.

Based on such measurements, we could estimate the system's response time at $30 \hspace{2pt} ms$, corresponding to modulation rates of 33 Hz. However, faster electronics and mechanical actuators, such as those used in deformable mirrors, can result in rates approaching the KHz regime \cite{archer2016dynamic}. Furthermore, since our approach to perturbing the fiber is general and not limited to mechanical perturbations, it could be directly transferred to other types of actuators, e.g., in-fiber acousto-optical or electro-optical modulators, which could reach even higher rates.

\subsection{Decorrelation Time}
Optical fibers are known to suffer from loss of correlation due to environmental fluctuations, such as thermal drifts and changes in air pressure. Bare fibers, such as the ones used with the fiber piano, are especially affected by these phenomena, as they do not have a protective outer jacket that isolates them from the surroundings. These effects lessen the system’s stability and are sometimes minimized by separating it from its surroundings, e.g. using a closed box in which the setup is placed. Our experiments ran in climate-controlled rooms, yet we did not take special care to isolate the system from the environment. 

To estimate the system's stability, we calculated the 2D Pearson correlation coefficient of the speckle pattern at the distal end of the fiber over time when the system is idle, i.e. no changes are performed to the states of the actuators. When beginning the measurement after the system was at rest, the speckle pattern stayed stable for many tens of hours. 

When performing random bending configurations, and then beginning the stability test, we find a difference between GRIN and SI fibers. With a GRIN MMF, we found that the system remained highly correlated ($corr \geq 0.99$) for $\simeq10$ minutes. The correlation decreased slowly and linearly for 55 minutes, reaching $corr=0.976$. The correlation then decreased faster, reaching $corr=0.883$ after two hours. With an SI few-mode fiber, the system remained stable and highly correlated ($corr \geq 0.996$) over 15 hours.

\subsection{Hysteresis} \label{hysteresis}
Our implementation relies on mechanical perturbations to the fiber, which are realized by an array of piezoelectric actuators. These are low voltage bimorph plate-benders, with a maximal travel of hundreds of microns. Such benders are known to have hysteresis, which means that a given voltage configuration will not always give the exact same bend configuration. This phenomenon can, in principle, decrease the optimization efficiency. However, when performing optimizations using the fiber piano, we did not observe a significant effect that we could attribute to hysteresis. This can be the result of good hyper-parameter choice \cite{rahman2014particle} or due to the nature of the algorithms, which upon convergence searches for a better solution using small variations from previously detected extrema. 

Since the fiber piano is a complex system, it may be interesting to investigate in the future the possibility of more intricate hysteresis effects involving several actuators. For instance, when changing actuators $A$ and $B$ from voltages ($V_{A1}$,$V_{B1}$) to ($V_{A2}$,$V_{B2}$), will it matter which voltage was changed first? If such effects occur, they will probably be between neighboring actuators since the actuators are placed relatively far from each other and the fiber is rather strongly held by each actuator even when no voltage is applied.

\section{Optimization parameters}
The optimization algorithm we chose in \cite{resisi2020wavefront, shekel2023shaping} is particle swarm optimization (PSO). An open-source implementation of PSO \cite{Yarpiz} was modified to fit our experimental setup and simulation. The PSO hyper-parameters which were used in our experiments were: inertia weight of $w=1$, inertia damping ratio of $w_{damp}=0.97-0.99$, personal learning coefficient of $c_1=1.5$, global learning coefficient of $c_2=2-2.2$. The population size used varied between experiments, between $40$ and $120$. 

The optimization algorithm we chose in \cite{finkelstein2023spectral} is the simulated annealing (SA) algorithm \cite{rutenbar1989simulated}. The optimization parameters we used are an initial temperature T of 20, with a cooling function that scales as T$\alpha^i$, where $i$ is the iteration number and $\alpha=0.97-0.99$. To decrease the probability of converging to local minima, at each iteration we add with some probability $p$ a weak perturbation to the optimal configuration of actuators and increase $p$ as the temperature decreases.

\printbibliography